# Enhancing quantum entropy in vacuum-based quantum random number generator

Xiaomin Guo[1,2], Ripeng Liu[1,2], Pu Li[1,2], Chen Cheng[1,2], Mingchuan Wu[1,2], and Yanqiang Guo[1,2]*

[1]*Key Laboratory of Advanced Transducers and Intelligent Control System, Ministry of Education and Shanxi Province, Taiyuan University of Technology, Taiyuan 030024, China*

[2]*Institute of Optoelectronic Engineering, College of Physics and Optoelectronics, Taiyuan University of Technology, Taiyuan 030024, China*

*E-mail: guoyanqiang@tyut.edu.cn

**Abstract:** Information-theoretically provable unique true random numbers, which cannot be correlated or controlled by an attacker, can be generated based on quantum measurement of vacuum state and universal-hashing randomness extraction. Quantum entropy in the measurements decides the quality and security of the random number generator. At the same time, it directly determine the extraction ratio of true randomness from the raw data, in other words, it affects quantum random numbers generating rate obviously. In this work, considering the effects of classical noise, the best way to enhance quantum entropy in the vacuum-based quantum random number generator is explored in the optimum dynamical analog-digital converter (ADC) range scenario. The influence of classical noise excursion, which may be intrinsic to a system or deliberately induced by an eavesdropper, on the quantum entropy is derived. We propose enhancing local oscillator intensity rather than electrical gain for noise-independent amplification of quadrature fluctuation of vacuum state. Abundant quantum entropy is extractable from the raw data even when classical noise excursion is large. Experimentally, an extraction ratio of true randomness of 85.3% is achieved by finite enhancement of the local oscillator power when classical noise excursions of the raw data is obvious.

## 1. Introduction

Randomness is one vital ingredient in modern information science, both classical and quantum [1,2], since encryption is founded upon the trust in random numbers [3-5]. The demand for true and unique randomness in these applications has triggered multitudinous proposals for producing random numbers based on measurements of quantum observables, which offer the verifiability and ultimate in randomness. In the past two decades, there has been tremendous development for various types of quantum random number generator (QRNG) [6-15]. Among these proposals, random number generation based on homodyne measurement of quantum vacuum state is especially appealing in practice since highly efficient photodiodes working at room temperature can be applied[11]. Vacuum state is the lowest energy pure quantum state and independent of any external physical quantities. It cannot be correlated or controlled by an attacker, therefore unique random numbers can be yielded by measuring the quadrature amplitude of the vacuum state [16,17]. All the components in this scheme, including laser source, beam splitter and photo detectors have been integrated on a single chip recently [18]. Meanwhile, bit conversion and post-processing are easy to implement in virtual "hardware" inside the field-programmable gate array (FPGA). Chip-size integration of the QRNG is expectable. Several dedicated researches have been developed in order to enhance the generation rate in this proposal, such as schemes based on optimization of the digitization algorithm [19], implementation of fast randomness extraction in the post-processing [20], squeezing vacuum state to increase entropy in raw data [21], and optimizing quantum entropy in raw data by optimized ADC parameters [22]. In this paper, considering the effects of the classical noise, we discuss the role of homodyne gain in enhancing quantum entropy in the vacuum-based QRNG in the optimum dynamical ADC range scenario. Conditional min-entropy is applied to critically assess the quantum entropy in the QRNG. It is the key input parameter of random extractor and determines the extraction ratio of true randomness from raw random sequence, thereby affects the generation speed of QRNG significantly.

## 2. Quantum entropy evaluation and enhancing in vacuum-based QRNG system

In optical homodyne tomography (OHT), balanced homodyne detection (BHD) system is established and locked to every relative phase to measure the marginal distributions of electromagnetic field quadrature for completely reconstruction of quantum states [23]. While the random numbers generation scheme discussed here focus on a marginal distribution of vacuum state in any one phase thanks to the space rotational invariance of its distributions in the phase space, i.e. no active modulation or phase (or polarization) stabilization is required. According to postulates of quantum mechanics, the measured $X$ quadrature of the vacuum states are totally random and satisfy the Gaussian distribution statistically, so we can extract random bits from the measurement results. In homodyne measurement, without regard to classical noise, the electric signal (voltage or current) is:

$$P(V_{vac}) = \frac{1}{\sqrt{\pi}\alpha} \exp(-\frac{V_{vac}^2}{\alpha^2}) . \tag{1}$$

The coefficient $\alpha$ has to be calibrated to rescale histogram of the associated marginal distribution in OHT. In the quantum random numbers generating proposal, $\alpha$ is associated with the quantum entropy contained in the measured data and it is the critical parameter for digitization of the measured analogue signal.

When classical noise, such as electronic noise and local noise due to imperfect balancing in BHD, is considered, the observed probability distribution of the electric signal is a convolution of the scaled vacuum state marginal distribution and the classical noise histogram

$$P_{obs}(V) = \frac{1}{\alpha} \int P(\frac{V'}{\alpha}) P_{cl}(V - V') dV' . \tag{2}$$

Without loss of generality, the broadband electric noise and the LO noise distribution can be assumed Gaussian:

$$P_{obs}(V) = \frac{1}{\sqrt{\pi}\sqrt{\alpha^2 + \beta}} \exp(-\frac{V^2}{\alpha^2 + \beta}) . \tag{3}$$

The vacuum noise and the classical noise are independent random variables that are normally distributed, thus their sum is also normally distributed with a total variance equal to the sum of the two variances.

According to Eqns. (1), (2), and (3), the vacuum state measurement yields a distribution as follows

$$P_{obs}(V) = \frac{1}{\sqrt{\pi}\sqrt{\alpha^2 + \beta}} \exp(-\frac{V^2}{\alpha^2 + \beta}) , \tag{4}$$

with the measurement variance

$$\sigma_{obs}^2 = \sigma_{quan}^2 + \sigma_{cl}^2 = (\alpha^2 + \beta)/2, \qquad (5)$$

where 2 is added to retain the normalization of the distribution. Then the quantum and classical noise ratio (*QCNR*) in the homodyne measurement system is defined as

$$QCNR = 10\,\mathrm{Lg}(\sigma_{quan}^2/\sigma_{cl}^2). \qquad (6)$$

The *QCNR* is quite related to the signal-to-noise ratio of homodyne detection, which is defined as the ratio between the mean square noise of the measured vacuum state and the electronic noise, i.e., the quantity

$$S = (\alpha^2 + \beta)/\beta = \sigma_{obs}^2/\sigma_{cl}^2, \qquad (7)$$

or the clearance between the shot noise power spectrum and electronic noise power in dB units, $10\,\mathrm{Log}(S)\,\mathrm{dB}$, reading on spectrum analyser. In other words, when the homodyne detection system works in linear region, the *QCNR* of the raw data can be forecasted from the clearance shown by spectrum analyser.

In this proposal of QRNG, as a continuous-variable, the measurement output consisting of scaled *X* quadrature of the vacuum state and the classical noise is discretized by an *n*-bit ADC with a dynamical range $[-R+\delta/2, R-3\delta/2]$. The sampled signals are binned over $2^n$ bins with width $\delta = R/2^{n-1}$ and assigned a corresponding bit combination with length of *n*. 3-bit ADC binning is shown in Figure 1(a) as an example.

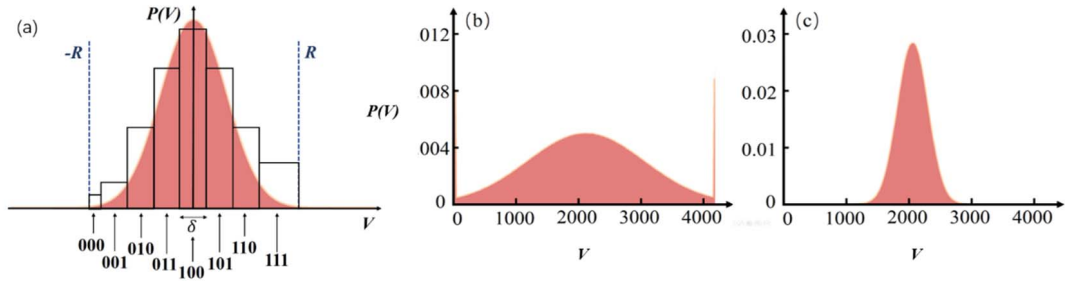

**Figure 1.** Model of 3-bit ADC (**a**) and Numerical simulations of acquisition conditions for a Gaussian signal when dynamical ADCZ range is chosen (**b**) too small and (**c**) too big.

The randomness or the entropy in the measurements may be coming from multiple factors including the quantum fluctuation, classical influences on it and even malicious attack from the third part [5]. The concept of entropy is applied as a count associated with the description. Especially, quantum conditional min-entropy is used to evaluate the maximal amount of randomness we can extract from the total entropy of the system [24] Firstly, the min-entropy for the Gaussian distribution is defined as

$$H_{\min}(X) = -\mathrm{Log}_2(\max_{V \in \{0,1\}^n} \mathrm{Prob}[X=V]). \qquad (8)$$

In this scheme, the min-entropy of the probability distribution *X* on $\{0,1\}^n$ can be accurately derived from the probability density function of the quantum signal. The maximum probability in (8) can be acquired based on the probability distribution discretized by the bins

$$P_{\text{bin}}(V_i) = \begin{cases} \int_{-\infty}^{-R+\delta/2} P_{\text{obs}}(V)dV, i = i_L, \\ \int_{V_i-\delta/2}^{V_i+\delta/2} P_{\text{obs}}(V)dV, i_L < i < i_M, \\ \int_{R-3\delta/2}^{+\infty} P_{\text{obs}}(V)dV, i = i_M. \end{cases} \quad (9)$$

Each bin is labelled by an integer $i \in \{i_L, ..., i_M\}$, with $i_L = -2^{n-1}$, the LSB bin, $i_M = 2^{n-1} - 1$, the MSB bin, and $V_i = i \times \delta$.

Secondly, some restrictions must be taken into account in analog-digital conversion process. Those samples go off-scale, i.e. points in saturation will be recorded as extrema values as depicted in Figure 1(b). So underestimating the range will induce too many blocks of zeros and ones. Conversely, overestimating the signal range will lead to undue unused bins (Figure 1(c)). Both these situations will cause some bit combinations too frequent to be considered random. It is necessary to adjust the amplitude of the analogue signal and the ADC dynamical range to use the full n-bit range whenever possible.

Further, considering the influence of classical noise on the measurement outcome, ADC dynamical range should be optimized over the classical noise shifted quantum signal probability distribution. In application scenario, inevitable classical noise excursion in the measurement system will result in nonzero mean in the measured signal probability distribution. On the other hand, eavesdropper may induce a deliberate offset over the sampling period. In a word, a noticeable classical noise excursion, $\Delta$, need to be considered in the optimization of the sampling dynamical range.

In this model, we rewrite the discretized probability distribution as,

$$P_{\text{bin}}(V_i|V_{\text{cl}}) = \begin{cases} \int_{-\infty}^{-R+\delta/2-\Delta} P(V_i|V_{\text{cl}})dX, i = i_L, \\ \int_{V_i-\delta/2-\Delta}^{V_i+\delta/2-\Delta} P(V_i|V_{\text{cl}})dX, i_L < i < i_M, \\ \int_{R-3\delta/2-\Delta}^{+\infty} P(V_i|V_{\text{cl}})dX, i = i_M. \end{cases} \quad (10)$$

In which,

$$P(V_i|V_{\text{cl}}) = \frac{1}{\sqrt{\pi}\alpha} \exp\left(-\frac{(V-V_{\text{cl}})^2}{\alpha^2}\right) \quad (11)$$

is the probability density distribution of the quantum signal given full knowledge of the classical noise $V_{\text{cl}}$, where $V_{\text{cl}} \in [V_{\text{cl,min}}, V_{\text{cl,max}}]$ with an excursion of $\Delta$. Finally, the quantum conditional min-entropy is expressed as

$$H_{\min}(V|V_{\text{cl}}) = -\text{Log}_2[\text{Max}(\frac{1}{2}\{1+\text{Erf}[\frac{-2(V_{\text{cl,min}}+R+\Delta)+\delta}{2\alpha}]\}, \\ \text{Erf}(\frac{\delta}{2\alpha}), \frac{1}{2}\{1+\text{Erf}[\frac{2(V_{\text{cl,max}}-R+\Delta)+3\delta}{2\alpha}]\})]. \quad (12)$$

In the best-case scenario of ADC sampling range, the measurement outcome probability in the centre bin is equal to the higher one of the first and the last bins. In this way, the quantum conditional min-entropy rigorously evaluates the amount of quantum-based randomness in the total noise signal. For applications that require information security, a random sequence is demanded to be truly unpredictable and have maximum entropy [25].

At the same time, the conditional min-entropy sets the lower bound of extractable randomness and quantifies the least amount of randomness possessed by each sample or $P = H_{\min}(X)/n$ bit per raw bit. Quantum randomness can be distilled from raw data by applying information theoretically provable Toeplitz-hash extractor. As discussed above, the key point is to find out the QCNR and derive the probability distribution of the quantum signal. The higher the QCNR, the more true randomness can be extracted from the raw measurement. Only when QCNR is high enough, both the quality and the security of the random number generator are guaranteed. Conditioned by optimal dynamical sampling range $R$, minimum-entropy of the quantum signal for growing clearance is theoretically analysed. Proceeding from the directly measurable quantity, homodyne clearance, corresponding QCNR is derived from equation (7). Then quantum noise variances are expressed as multiples of the $\sigma_{cl}$. For different clearance, probabilities of middle bin and the LSB/MSB are compared and the optimal sampling range $R$ is decided based on equation (10). Finally, based on equation (12), the quantum conditional min-entropy in optimal sampling range scenario as a function of different classical noise excursion is analysed.

The classical noise excursions in our raw data have been collected from multiple measurements, which range from almost 3 to 29 times of classical noise standard deviation $\sigma_{cl}$. In application scenario, much larger DC offset may be induced deliberately by the eavesdropper. In Figure 2, we show the quantum conditional min-entropy, $H_{\min}(V|V_{cl})$, as a function of homodyne detection clearance for three different classical noise excursion under the precondition of optimal sampling range. $\Delta = 3\sigma_{cl}$ is the smallest classical noise excursion among our multiple measurements, $\Delta = 40\sigma_{cl}$, a larger classical noise excursion for comparison, and $\Delta = 17.2\sigma_{cl}$ is the excursion in the raw data from which we extract true random numbers. As shown in Figure 2, the extractable random bits are robust when the classical excursion is subtle. Whereas if classical noise excursion is evidence, one can achieve high secure randomness only when clearance is high enough.

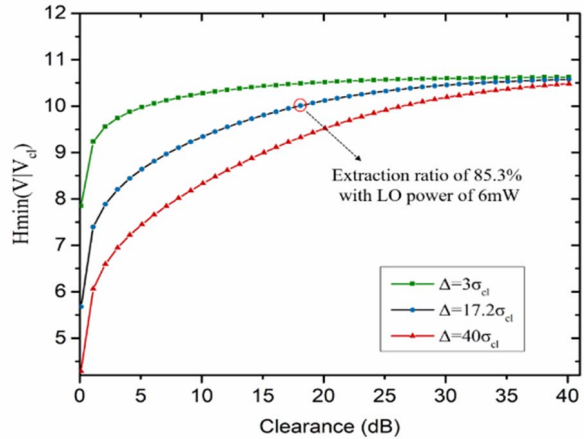

**Figure 2**. Optimized $H_{\min}(V|V_{cl})$ as a function of homodyne detection clearance among different classical excursions. The theoretical value circled in red corresponding to the highest extraction ratio of true randomness in our experiment.

The clearance lies on the total gain in homodyne detection system (also $\alpha$ in equation (1)), including the LO amplification and the electrical gain of transimpedance amplifier (TIA). In quantum state measurements and reconstructions, the clearance needed between shot noise and classical noise is dependent on the amount of squeezing and entanglement one wishes to measure. Empirically, the homodyne system should satisfy the condition that the measured shot noise is 10 dB higher than the classical noise among the analysis frequency range [26,27]. High TIA gain and moderate dynamical range are required so that shot noise is the dominant spectral feature among the detection frequency range. In this scheme of QRNG, however, high $QCNR$ but also large detection bandwidth are pursued, since cut-off frequency of the homodyne detector upper bound the sampling frequency in random numbers generation process [28].

On the other hand, the classical effects which blur the distribution and cause classical entropy in the raw bit sequence include imperfect balancing of LO, non-unit quantum efficiency, and electronic noise of the detectors [29–32]. The imperfection of detector efficiency can almost completely overcome by using special fabricated diodes, and the quantum efficiencies of more than 99% have been reported [33]. The detrimental electronic noise depends on numerous components in the circuit part as expressed by

$$V_{\mathrm{EL,noise}} = R\sqrt{(4KT/R_{\mathrm{PD}} + I_{\mathrm{PD,dk}}^2 + 4KT/R_{\mathrm{r}} + I_{\mathrm{TIA,c}}^2) + (V_{\mathrm{TIA,v}}/R)^2} \qquad (13)$$

One part is from the photodiode (PD) and comprise of thermal noise and dark current noise of PD, both of which are usually negligible thanks to its big shunt resistance $R_{\mathrm{PD}}$ and low dark current $I_{\mathrm{PD,dk}}$ [34]. The other part is from the TIA circuit including thermal noise $4KT/R_{\mathrm{r}}$, input noise current $I_{\mathrm{TIA,c}}$, and input noise voltage $V_{\mathrm{TIA,v}}$ of the operational amplifier. The electrical gain of TIA amplifies quantum fluctuations as well as the electronic fluctuations, so the electronic noise included in the homodyne raw measurements comes mainly from the amplified TIA circuit noise. Unlike that though, local oscillator effectively acts as a noise-less amplifier for the quantum fluctuations of the vacuum state and the electrical noise is independent of the LO. In fact, the optical fluctuations seen by the detector can be made much larger than the electronic fluctuations by enhancing the LO intensity to enhance the $QCNR$ signally [35].

At the same time, the gain of a typical op-amp is inversely proportional to frequency and characterized by its gain–bandwidth product (GBWP). As a trade-off, lower electrical gain put up with higher op-amp bandwidth. In fact, the bandwidth of the vacuum noise is infinite, the random number generation rate in this scheme is ultimately limited by the bandwidth of the homodyne detector. Increased bandwidth of op-amp allows higher sampling rate.

### 3. Experiment and results

Experimentally, we dedicate to enhance quantum entropy in QRNG from vacuum fluctuation by enhancing local oscillator intensity to noise-independent amplify quadrature fluctuation of vacuum state on the premise of optimize ADC sampling range. An extraction ratio of true randomness of 85.3% is achieved by finite enhancement of the local oscillator power when classical noise excursions of the raw data is obvious and the extracted random sequences passed the NIST, Diehard and the TestU01 tests.

The experimental setup is depicted in Figure 3. A 1550 nm laser diode (LD) is driven by constant current with thermoelectric cooling control with a maximal out power of 15mW. A half-wave plate and a polarizing beamsplitter (PBS2) were combined to realize 50/50 beamsplitting ratio. Single-mode continuous-wave laser beam from the laser incident into one port of the beamsplitter and acts as the LO, while the other port was physically blocked to ensure that only the vacuum state could enter in. The vacuum state and the LO interfere on the symmetric beamsplitter to form two output beams with balanced power. The outputs are simultaneously detected by balanced homodyne detector to cancel the excess noise in LO while amplify the quadrature amplitude of the vacuum, $X$, which fluctuates totally randomly and is independent of any external physical quantities thus cannot be tempered with.

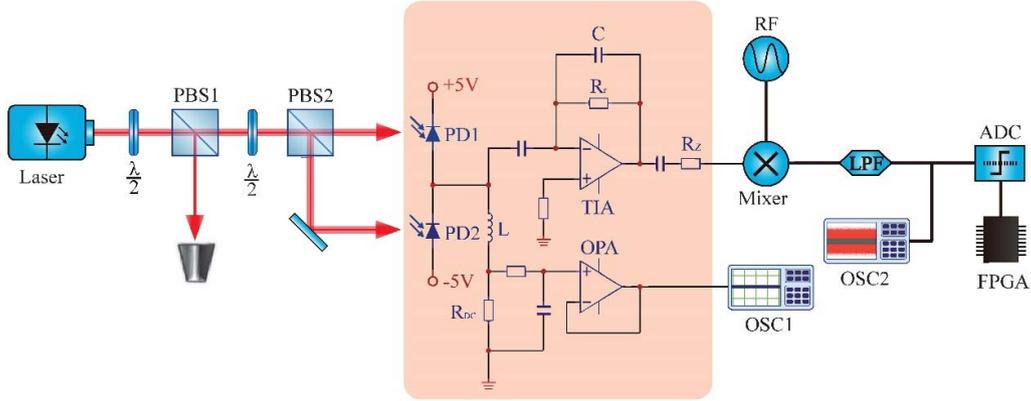

**Figure 3.** Schematic of the experiment for the quantum random number generator based on homodyne measurements of the quadrature amplitudes of the vacuum state.

Classical noise in the photocurrents is rejected effectively over the whole detection band while the clearance have dependence on frequency as shown in Figure. 4. We filtered out a part of the vacuum spectrum, where the clearance is almost consistent, to extract true randomness based on a certain quantum conditional min-entropy and analysis the effect of LO intensity on *QCNR*. The shot noise current outcome from the homodyne detector is mixed down with a 200MHz carrier (HP8648A) and pass through a low-pass-filter (LPF) with 50MHz cut-off frequency (Mini-Circuits BLP50+ ), that is, we actually use 100MHz vacuum sideband frequency spectrum centred at 200MHz to act as the random entropy resource.

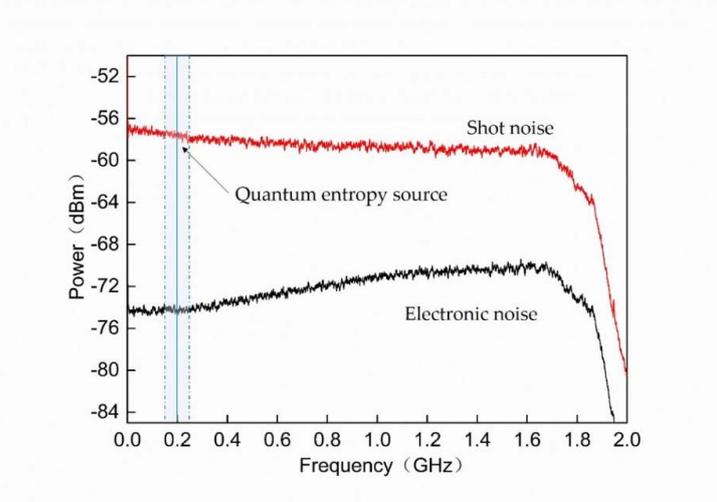

**Figure 4.** Amplified vacuum noise power spectral when LO power is 6mW. 100MHz vacuum sideband centred at 200MHz is filtered out as the entropy source of QRNG.

We present the *QCNR* as a function of the LO power arriving at the PD. The electrical noise variance is relatively consistent for certain TIA gain. The clearance depends only on the LO power. The noise power is given by

$$P_{dBm} = 10\lg(\frac{4e^2(P/h\nu)\eta BR^2}{Z \times 1mW}), \tag{14}$$

where $e$ is the electron charge, $\eta = 0.9$ is the quantum efficiency of the photodiode (Hamamatsu G8376), $B = 100\,\text{KHz}$ the resolution bandwidth, $R = 16 \times 10^3$ V/A the transimpedance gain of the photo detector and $Z = 50\Omega$ the load impedance [36]. For each power value the distribution of the random data was analysed in time domain and histogram to calculate the *QCNR*. *QCNR* as a function of the LO power figured out from the measured clearance levels is plotted with open circles in Figure 5. The LO power received by each PD is progressively increased from 300μW to 6mW by rotating the HWP before PBS1. Because the theory gives noise power as a function of LO power arriving at the PD, we interpolate between the experimental points to obtain the dependence of *QCNR* on LO power. It is shown as the black dashed line in Figure 5. The experimental results are given by red open circles and can be fitted well by the theoretical curve with a transimpedance gain of $13.1 \times 10^3$ V/A . This is due to uncertainties in determining the transimpedance of the detector and the transmission losses in the LPF.

We enhance the local oscillator power to 6mW to achieve the largest *QCNR* of 17.8dB in our system, limited by the maximal output power of the laser. The signal is sampled with a rate of 100MHz, upper limit of twice the LPF band for the sampling rate to avoid temporal correlation between samples. The resolution is 12 bit and the dynamical range is optimized according to the histogram of the time series acquired with reasonably larger sampling range. The amplitude acquisition scale continuous adjustable oscilloscope (LeCroy SDA806Zi-A) is used. By choosing the analog-digital conversion range appropriately and tuning the LO intensity finely, the amount of off-scale points can be controlled within allowed statistical deviation. The number of saturated points is easy to restrain on-line from the oscilloscope. The distributions of the random data in time domain and in histogram are shown as insets of Figure 5. The measured total variance of the raw data and electrical noise variance are $154.43\,\text{mV}^2$ and $5.89\,\text{mV}^2$, respectively. The classical noise excursions of the raw data is about 17.2 times of the classical noise standard deviation $\sigma_{\text{cl}}$. Then the probability distribution of the quantum signal is derived and the min-entropy in the quantum signal is worked out to be 10.13 bit per sample, as circled in red in Figure 2.

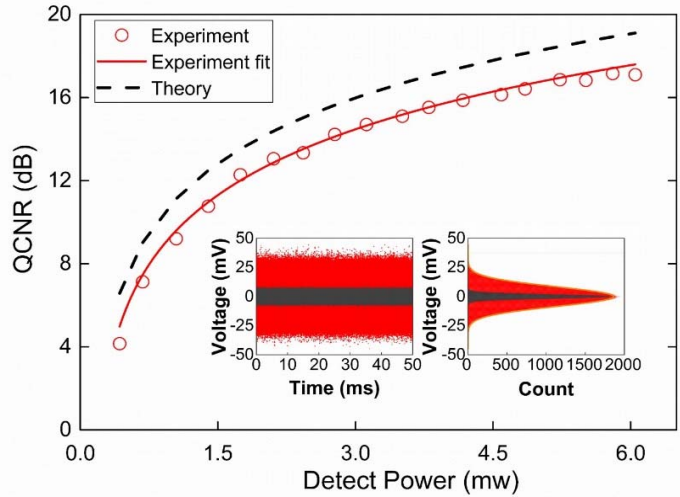

**Figure 5**. *QCNR* as a function of the LO power. Inset: Resulting histograms of the vacuum (red) and electronic (black) noise obtained at a LO power of 6mW.

Finally, information-theoretically provable post-processing scheme, Toeplitz-hashing extractor, is constructed on an FPGA to extract true randomness from the raw data and uniform the Gaussian biased binary stream [37]. A binary Toeplitz matrix of $m \times n$ is constructed with a seed of $m+n-1$ random bits (the seed can be reused since the Toeplitz-hashing extractor is a strong extractor). $m$ final random bits are extracted by multiplying the matrix and $n$ raw bits, where $m/n \le P$ and $P = H_{\min}(X)/n$. We employ $4096 \times 3520$ Toeplitz Hash extractor to distil 10.13 bits/sample. The extraction ratio of 85.3% is the highest as ever reported. We recorded a file with data size of 1G bits to undergo random test. 1000 sequences with each one 1Mbits are applied to the NIST test and significant level is set as $\alpha=0.01$ . The

NIST test is successful if final P-values of all sequences are larger than $\alpha$ with a proportion within the range of $(1-\alpha) \pm 3\sqrt{(1-\alpha)\alpha/n} = 0.99 \pm 0.00944$ for 15 test suits [38]. P-value shown in the Figure 6 are the worst cases of our test outcomes.

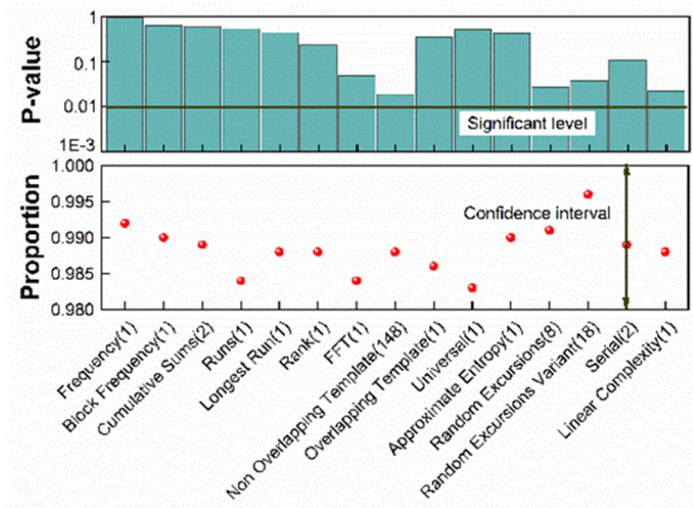

**Figure 6**. Results of the NIST statistical test suite for a $10^9$-bit sequence.

Results of the Diehard statistical test suite for the same data file is shown in Figure 7. Kolmogorov-Smirnov (KS) test is used to obtain a final p-value to measure the uniformity of the multiple P-values. The test is considered successful if all the final P-values lies in the range from 0.01 to 0.99 [39].

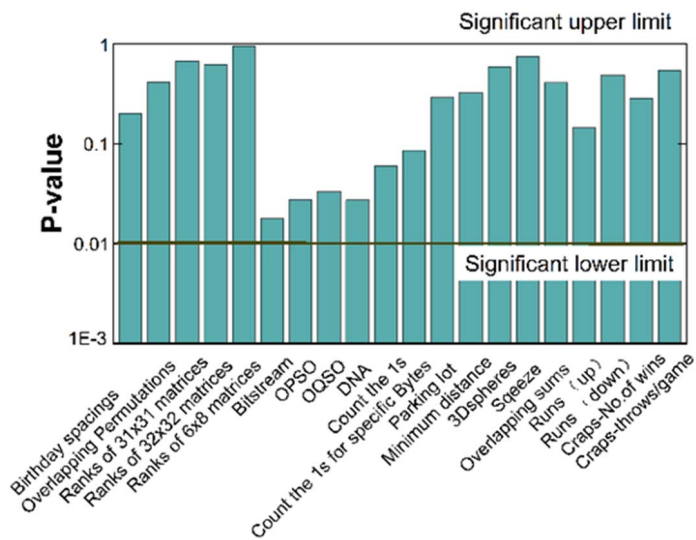

**Figure 7**. Results of the Diehard statistical test suite for a $10^9$-bit sequence.

Constrained by the computational power of crush of TestU01, small crush test is performed to a data file of 8 Gbits [40]. The random numbers can pass all the statistical tests successfully. The P value from a failing test converge to 0 or 1. Where the test has multiple P values, the worst case is tabled in Figure 8.

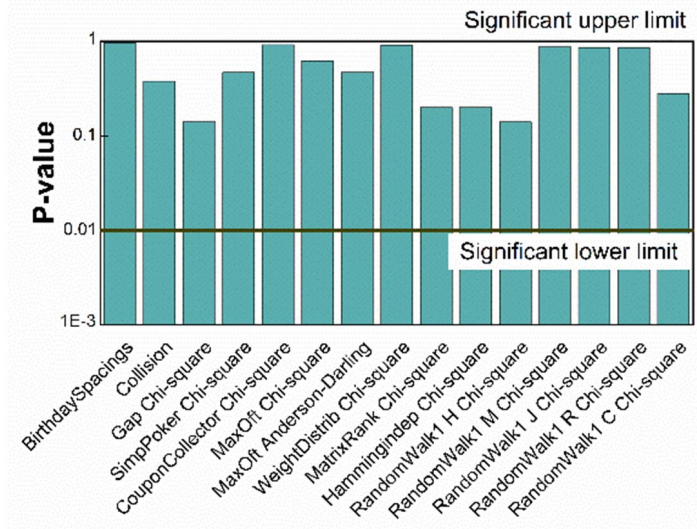

**Figure 8**. Results of the TestU01 statistical test suite for a $5\times10^9$-bit sequence.

On the other hand, we lower the LO power in the homodyne system to 400μW and thereby the clearance decline to 4.06dB. The time series of the system outcomes are collected and statistically analysed. Classical noise excursion in the Gaussian distribution is about 19.3 times of the classical noise standard deviation. Based on these results, the min-entropy of the measurements is worked out to be 7.73 bits/sample. The hash extraction result with an extraction ratio of 0.63 pass the NIST, Diehard and TestU01 tests finally.

### 4. Conclusions

To summary, in this work, we discussed the role of LO power plays in random number generation based on quantum detection of vacuum state. When classical noise excursion in the system is trivial, LO power in the homodyne system affect the quantum entropy in raw data insignificantly. Nevertheless, in realistic scenario, the mean of the measured signal distribution is normally nonzero, even much larger noise excursion may be induced deliberately by the eavesdropper. In this case, enough real randomness is attainable only when *QCNR* is high enough. With the LO power enhanced, the vacuum quadrature fluctuations is amplified independent of the electrical noise and the *QCNR* is enhanced effectively. Thus, we propose large dynamical range and moderate TIA gain to pursue higher LO amplification of vacuum quadrature and larger detection bandwidth in homodyne system for higher sampling rate in random numbers generation. Higher hash extraction ratio along with higher sampling rate will enhance the real random number generation rate effectively. More importantly, the QRNG system is more robust against to the third part attack.

**References**

[1] Korzh, B.; Lim, C. C. W.; Houlmann, R.; Gisin, N.; Li, M. J.; Nolan, D. Provably secure and practical quantum key distribution over 307km of optical fibre. *Nat. Photonics*. **2014**, *9*, 163-168.

[2] Ferguson, N.; Schneier, B.; Kohno, T. *Cryptography Engineering: Design Principles and Practical Applications*; Wiley Publishing, **2010**.

[3] Stefanov, A.; Gisin, N.; Guinnard, O.; Guinnard, L.; Zbinden, H. Optical quantum random number generator. *J. Mod. Optic*. **2000**, *47*, 595-598.

[4] Gisin, N.; Ribordy, G.; Tittel, W.; Zbinden, H. Quantum cryptography. *Rev. Mod. Phys*. **2002**, *74*,

145–195.

[5] Toffoli, T. Entropy? honest!. *Entropy*. **2016**, *18*, 247.

[6] Rarity, J.; Owens, P.; Tapster, P. Quantum Random-number Generation and Key Sharing. *Optica Acta International Journal of Optics*. **1994**, *41*, 2435-2444.

[7] Guo, H.; Tang, W. Z.; Liu, Y.; Wei, W. Truly random number generation based on measurement of phase noise of a laser. *Phys. Rev. E*. **2010**, *81*, 051137.

[8] Ma, H. Q.; Xie, Y.; Wu, L. A. Random number generation based on the time of arrival of single photons. *Appl. Opt.* **2005**, *44*, 7760-7763.

[9] Yan, Q. R.; Zhao, B. S.; Liao, Q. H.; Zhou, N. R. Multi-bit quantum random number generation by measuring positions of arrival photons. *Rev. Sci. Instrum.* **2014**, *85*, 615-621.

[10] Ren, M.; Wu, E.; Liang, Y.; Jian, Y.; Wu, G.; Zeng, H. P. Quantum random-number generator based on a photon-number-resolving detector. *Phys. Rev. A*. **2011**, *83*, 1293-1304.

[11] Gabriel, C.; Wittmann, C.; Sych, D.; Dong, R. F, Mauerer, W.; Andersen, U. L. A generator for unique quantum random numbers based on vacuum states. *Nat. Photonics.* **2010**, *4*, 711-715.

[12] Qi, B.; Chi, Y. M.; Lo, H.-K.; Qian, L. High-speed quantum random number generation by measuring phase noise of a single-mode laser. *Opt. Lett.* **2010**, *35*, 312–314.

[13] Xu, F. H.; Qi, B.; Ma, X. F.; Xu, H.; Zheng, H. X.; Lo, H. K. Ultrafast quantum random number generation based on quantum phase fluctuations. *Opt. Express*. **2012**, *20*, 12366.

[14] Marangon, D. G.; Vallone, G.; Villoresi, P. Source-device-independent ultrafast quantum random number generation. *Phys. Rev. Lett.* **2017**, *118*, 060503.

[15] Cao, Z.; Zhou, H.; Ma, X. F. Loss-tolerant measurement-device-independent quantum random number generation. *New. J. Phys.* **2015**, *17*, 125011.

[16] Sych, D; Leuchs, G. Quantum uniqueness, *Found Phys.* **2015,** *45,*1613–1619.

[17] Fiorentino, M.; Santori, C.; Spillane, S. M.; Beausoleil, R. G.; Munro, W. J. Secure self-calibrating quantum random-bit generator. *Phys. Rev. A*. **2006**, *75*, 723-727.

[18] Abellan, C.; Amaya, W.; Domenech, D.; Muñoz, P.; Capmany, J.; Longhi, S. Quantum entropy source on an InP photonic integrated circuit for random number generation. *Optica*. **2016**, *3*, 989-994.

[19] Symul, T.; Assad, S. M.; Lam, P. K.. Real time demonstration of high bitrate quantum random number generation with coherent laser light. *Appl. Phys. Lett.* **2011**, *98*, 231103.

[20] Shi, Y. C.; Chng, B.; Kurtsiefer, C. Random numbers from vacuum fluctuations. *Appl. Phys. Lett.* **2016**, *109*, 041101.

[21] Zhu, Y. Y.; He, G. Q.; Zeng, G. H. Unbiased quantum random number generation based on squeezed vacuum state. *Int. J. Quantum. Inf.* **2012**, *10*, 1250012.

[22] Haw, J. Y.; Assad, S. M.; Lance, A. M.; Ng, N. H. Y.; Sharma, V.; Lam, P. K. Maximization of extractable randomness in a quantum random-number generator. *Phys. Rev. Appl.* **2015**, *3*.054004.

[23] Lvovsky, A. I.; Raymer, M. G. Continuous-variable optical quantum state tomography. *Rev. Mod. Phys.* **2005**, *81*, 299–332.

[24] Konig, R.; Renner, R.; Schaffner, C. The operational meaning of min- and max-entropy. *IEEE T. on Inform. Theory* **2009**, *55*, 4337-4347.

[25] Stipčević, M. Quantum random number generators and their applications in cryptography.

*Advanced Photon Counting Techniques VI*, **2012**, *8375*, 837504.

[26] Vahlbruch, H.; Mehmet, M.; Chelkowski, S.; Hage, B.; Franzen, A.; Lastzka, N. Observation of squeezed light with 10-db quantum-noise reduction. *Phys. Rev. Lett.* **2008**, *100*, 033602.

[27] Olivares, S.; Paris, M. G. A. Bayesian estimation in homodyne interferometry. *J. Phys. B: At. Mol. Opt. Phys.* **2012**, *42*, 55506-55512.

[28] Shen, Y.; Tian, L.; Zou, H. X. Practical quantum random number generator based on measuring the shot noise of vacuum states. *Phys. Rev. A* **2010**, *81*, 063814.

[29] Mcclelland, D. E.; Mckenzie, K.; Gray, M. B.; Ping, K. L. Technical limitations to homodyne detection at audio frequencies. *Appl. Optics.* **2007**, *46*, 3389-3395.

[30] Gramdi, S.; Zavatta, A.; Bellini, M.; Paris, M. G. A. Experimental quantum tomography of a homodyne detector. *New. J. Phys.* **2017**, *19*, 053051.

[31] Combes, J.; Wiseman, H. Quantum feedback for rapid state preparation in the presence of control imperfections *J. Phys. B: At. Mol. Opt. Phys.* **2011**, *44* 154008.

[32] Chrzanowski, H. M.; Assad, S. M.; Bernu, J.; Hage, B.; Lund, A. P.; Ralph, T. C. Reconstruction of photon number conditioned states using phase randomized homodyne measurements. *J. Phys. B: At. Mol. Opt. Phys.* **2013**, *46*, 104009.

[33] Oshima, T.; Okuno, T.; Arai, N.; Suzuki, N.; Ohira, S.; Fujita, S. Vertical solar-blind deep-ultraviolet schottky photodetectors based on beta-$Ga_2O_3$ substrates. *Appl. Phys. Express.* **2008**, *1*, 011202.

[34] Graeme, J. *Photodiode Amplifiers: OP AMP Solutions*; McGraw-Hill: New York, 1995.

[35] Jin, X. L.; Su, J.; Zheng, Y. H.; Chen, C.; Wang, W. Z.; Peng, K. C. Balanced homodyne detection with high common mode rejection ratio based on parameter compensation of two arbitrary photodiodes. *Opt. Express.* **2015**, *23*, 23859.

[36] Gray, M. B.; Shaddock, D. A.; Harb, C. C.; Bachor, H.-A. Photodetector designs for low-noise, broadband, and high-power applications. *Rev. Sci. Instrum.* **1998**, *69*, 3755-3762.

[37] Carter, J. L.; Wegman, M. N. Universal classes of hash functions (Extended Abstract). *J. Comput. Syst. Sci.* **1977**, *18*, 106-112.

[38] Rukhin, A.; Soto, J.; Nechvatal, J.; Miles, S.; Barker, E.; Leigh, S. A statistical test suite for random and pseudorandom number generators for cryptographic applications. *Appl. Phys. Lett.* **2015**, *22*, 1645-179.

[39] Marsaglia G *"DIEHARD: A battery of tests of randomness."* 1996, Online: Available: http://www.stat.fsu.edu/pub/diehard/.

[40] L'Ecuyer, P.; Simard, R. TestU01: A C library for empirical testing of random number generators. *ACM*. **2007**, *33*, 22.